\documentclass[letter]{aa} 

\usepackage{graphicx}
\usepackage{txfonts}
\begin{document}

   \title{Extreme adaptive optics astrometry of R136\thanks{Catalogue of reliable-consistent sources is only available in electronic form
at the CDS via {\protect \url{http://cdsweb.u-strasbg.fr/cgi-bin/qcat?J/A+A/}} }}
  \subtitle{Searching for high proper motion stars\thanks{Based on observations made with ESO Telescopes at the La Silla Paranal Observatory under programme ID 0102.D0271 and 095.D-0309} }


   \author{Z. Khorrami
          \inst{1}
          \and
          M. Langlois\inst{2}
          \and
          F. Vakili\inst{3}
          \and
          P. C. Clark\inst{1}
          \and
          A. S. M. Buckner\inst{4}
          \and
          M. Gonzalez\inst{5}
          \and
          P. Crowther\inst{6}
          \and
          R. W\"unsch\inst{7}
          \and
          J. Palou\v{s}\inst{7}
          \and
          A. Boccaletti\inst{8}
          \and
          S. Lumsden\inst{9}
          \and
          E. Moraux\inst{5}
          }

   \institute{School of Physics and Astronomy, Cardiff University, The Parade, Cardiff CF24 3AA, UK\\
              \email{KhorramiZ@cardiff.ac.uk}
         \and
             Universite de Lyon, Universite Lyon 1, CNRS, CRAL UMR5574, Saint-Genis Laval, France
        \and
        Universite Cote d'Azur, OCA, CNRS, Lagrange, France
        \and
        School of Physics and Astronomy, University of Exeter, Stocker Road, Exeter, EX4 4QL, UK
        \and
        Universite Grenoble Alpes, CNRS, IPAG, F-38000 Grenoble, France
        \and
        Department of Physics and Astronomy, Hounsfield Road, University of Sheffield, Sheffield, S3 7RH, UK
        \and
        Astronomical Institute of the Czech Academy of Sciences, Bo\v{c}n\'\i II 1401/1a, 141 00 Praha 4, Czech Republic
        \and
        LESIA, Observatoire de Paris, CNRS, Universite Paris 7, Universite Paris 6, 5 place Jules Janssen, 92190 Meudon, France
        \and
        School of Physics and Astronomy, University of Leeds, Leeds LS2 9JT, UK
             }

   \date{}

 
  \abstract
   {We compared high-contrast near-infrared images of the core of R136 taken by VLT/SPHERE, in two epochs separated by 3.06 years.
   For the first time we monitored the dynamics of the detected sources in the core of R136 from a ground-based telescope with adaptive optics.
   The aim of these observations was to search for High prOper Motion cAndidates (HOMAs) in the central region of R136 ($r<6"$) where it has been challenging for other instruments.
   Two bright sources (K$<15$~mag and V$<16$~mag) are located near R136a1 and R136c (massive WR stars) and have been identified as potential HOMAs.
   These sources have significantly shifted in the images with respect to the mean shift of all reliable detected sources and their neighbours, and six times their own astrometric errors. We calculate their proper motions to be $1.36\pm0.22$~mas/yr ($321\pm52$~km/s) and $1.15\pm0.11$~mas/yr ($273\pm26$~km/s).
  We discuss different possible scenarios to explain the magnitude of such extreme proper motions, and argue for the necessity to conduct future observations to conclude on the nature of HOMAs in the core of R136.
  }


\maketitle
%

\section{Introduction}

The R136 star cluster in the 30\,Dor region of the Large Magellanic Cloud (LMC) is a unique astrophysical laboratory;  it is sufficiently young (1.5-2~Myr, \citealt{dekoter1998,crowther2016,Bestenlehner2020}) and rich \citep{crowther2010} to allow the  study of the formation and evolution of the most massive stars. 
Most of the  information on R136  comes from Hubble Space Telescope (HST) observations, which was used to study various parameters including the cluster’s star formation history, stellar content, and kinematics \citep{hunter95,andersen2009, demarchi2011,Cignoni2015}.  
Using the Spectro-Polarimetric High-contrast Exoplanet Research (SPHERE, \citealt{sphere}) of the Very Large Telescope (VLT) in 2015 we obtained the first epoch observations of the R136  core \citep{khorrami2017} in the near-infrared (NIR), and in 2018 we obtained a second set of observations for this region \citep{khorrami2021}.
Comparing the two sets of observations, we monitored the dynamics of the detected sources in the core of R136 searching for High prOper Motion cAndidates (HOMAs\footnote{Homa is a mythical bird from Iranian legends that never comes to rest, living its entire life flying invisibly high above the earth.}).
The presence of such high-velocity stars in the core of massive young star clusters can explain or change the hypothesis on the formation of massive stars (if they form either in isolation or were ejected dynamically from the cluster) and the dynamics of the 30\,Dor region as a whole. 

The proper motion (PM) measurement of stars in 30\,Dor relies on HST data taken with various instruments (WFPC2, ACS/WFC, and WFC3/UVIS) and filters in different epochs \citep{platais2015,platais2018}. There are therefor several difficulties with obtaining these measurements since the data in each instrument (and filter) show coordinate-dependent systematic and geometric distortions that appear in the transformation of instrument pixel coordinates and the PMs. 
Among the higher-precision PM sources, a handful of stars exceed 1mas/yr, but   were excluded from the analysis by setting an upper limit on the velocity of stars based on the maximum radial velocity measurements from VLT Flames Tarantula Survey (VFTS) observations \citep{evans2011}.
The PM measurements in the 30\,Dor region, as given by HST and recently by Gaia \citep{lennon2018}, do not detect sources at the core of R136 because of the concentration of bright sources and their small angular separations.
Thus, higher angular resolution and small pixel sampling are needed to improve the PM measurements without overexposing the bright sources (and masking their faint neighbours);   better estimations of the positions of stars are also needed during the photometry for point spread function (PSF) fitting.
SPHERE/IRDIS can observe the core of R136 in the NIR where the flux ratio of stars is lower than in the optical, and with higher angular resolution and pixel sampling (12.25~mas/pix) than the HST (e.g. WFC3/UVIS has 40~mas/pix).
These parameters make IRDIS the best instrument available at present to resolve the higher number of stars in the core of R136. 
Additionally, IRDIS produces two sets of data simultaneously taken within the same filter, overcoming the instrumental errors such as bad pixels and imperfections of the detector, which are critical to the PM measurements.

\section{Data} \label{sec:data}

We  used two sets of imaging data over a field of view (FOV) of 11"$\times$ 12" centred on the core of R136 in the K band in 2015 (2015-09-22, ID \url{095.D-0309}) and 2018 (2018-10-10, ID \url{0102.D-0271}). 
These two sets of epoch data were recorded using the classical imaging mode of IRDIS \citep{maud14}. Using the data taken with the same instrument and exactly the same configuration, 3.06 yr apart, enabled us to study the dynamics of the core of R136 with the highest angular resolution to date.
For our purposes, the same spectral band was split into the two IRDIS channels (the simultaneous images on the left and right side of
the detector), which were used to correct for residual detector artefacts such as hot pixels and uncorrelated detector noise, among other instrumental effects. 
We corrected  each image for the anamorphism following \cite{maire} before the final image combination.
The image sharpness of the total exposure was maximised by discarding the single frames with poorer Strehl ratio and by correcting a posteriori the residual tip-tilt image motion on each short exposure before combining them. 

The seeing was $0.69 \pm 0.10"$ and $0.63 \pm 0.10$, for the 2018 and 2015 observations, respectively. 
The night was rated as clear for both epochs, meaning that less than 10\% of the sky (above 30 degrees elevation) was covered by clouds, and transparency variations were less than 10\% during the exposures, with an airmass between 1.61-1.67 (in 2015) and 1.52-1.45 (in 2018).
Our data consists of 300 (in 2015) and 544 (in 2018) frames of 4.0s exposures taken with the IRDIS broad-band K filter ($\lambda_{cen} = 2182$~nm, $\Delta\lambda = 294$~nm).
The Wolf-Rayet star R136a1 was used for guiding the AO loop of SPHERE confirming the high level of performance even for faint guide stars, which surpasses  the performance of the Nasmyth Adaptive Optics System Near-Infrared Imager and Spectrograph (NACO) and the Multi-conjugate Adaptive optics Demonstrator (MAD)  (e.g. \citealt{campbell2010}).
The details of the data reduction procedure are given in Sect. 2 of \cite{khorrami2021}.

\section{Photometry}\label{sec:photometry}
Photometry is carried out on the two images (left and right)  from each epoch, using the same method and criteria as explained in Sect. 3 of \cite{khorrami2021}.
We needed to repeat the photometric analysis in both epochs  for three reasons:
    1) the input PSF should be the same in two epochs in order to have a robust photometric analysis;
    2) some of the input PSF used in the photometric analysis in the first epoch \citep{khorrami2017} appeared as multiple systems in 2018 where we use longer exposures; and
    3) the photometric analysis of the 2015 data was done on the combined left and right images, but  separate photometric analyses of the left and right images  were required for the present purpose.
    
The common sources between the two images were defined according to their position \footnote{Closer than the full width at half maximum (FWHM) of the PSF, which is the resolution}, their magnitude difference being less than 0.5, and the correlation between their input PSF and detected sources less than 10\%. This means that the detected sources in the two images have a very similar PSF shape since the input PSF stars are chosen to be the same in all four images.
The number of  common sources between the  left and right K images was 705 and 1451 in 2015 and 2018, respectively. 
Although in principle we should not see any shift in the position of the common sources per epoch, our astrometry analysis still measures the shift for each detected source in each epoch between the  left and right images. This shift can be caused by the instrument, detector bad pixels, locally varying pixel scales, and flux difference between the  left and right images. We call this the  ``instrumental shift'' because it does not represent potential physical movement of the stars.
If it is larger than half a pixel, we withdraw the source from our final astrometry analysis.
After applying this constraint, a total of  515 (in 2015) and 1059 (in 2018) sources in common between the  left and right images remain. 
The final number of common sources after this selection is 494. 

\section{Astrometry}\label{sec:astrometry}
We analysed the movement (shift) of each source locally, by superimposing its neighbours located within the radius of 50 pix. If there are fewer than  three neighbours, a larger radius (previous radius plus 10 pix) is  considered. 
A bootstrap technique is employed to measure the errors.
The astrometry is performed using a three-step process in order to superimpose data between two epochs:
1) local correction technique (LCT) to superimpose the data as much as possible to find any x-y shifts or rotation of data between the two epochs;
2) global rotation around a centre and angle found in the first step; and 
3) LCT again to correct any residual instrumental or artificial effects.

We used these steps in order to detect HOMAs in our K-band data between 2015 and 2018. These measurements were performed separately on the right and left data.
\subsection{Local correction and global rotation techniques}

In order to estimate the shift of stars, we used N detected neighbour sources within a radius of $R_{nbr}=50 pix$ in both images with the aim of superimposing these neighbour sources (similar to the method by \citealt{lct2010}). 
Our algorithm estimates the shift between the position of a given neighbour star in the 2015 and 2018 images in the X and Y directions ($dX_i$ and $dY_i$). By minimising the chi-square ($\chi^2$) estimation, the final $dX$ and $dY$ values are calculated and a bootstrapping technique ($N_{sample}=100$) is used to calculate the error on the $dX$ and $dY$ values. 
In the ideal case, without local distortion or bad pixels on the image (detector), the $dX$ and $dY$ values should be constant across the FOV so we can superimpose the two sets of  data easily by using the constant values of $dX$ and $dY$. 
However, as this does not happen in real data, we used an algorithm to superimpose the catalogues locally.

The variation in $dX$ and $dY$ across the FOV shows the global rotation centred at [634,669] and [622,932] in the left and right images in 2015, about $0.18^{o}$ clockwise.
The two sets of
catalogues were superimposed using this global rotation, and LCT was run again to remove any astrometric residuals and measure the PM of detected sources with higher precision.
The measured PM of the stars within 3.06 years are

$\mu_{x}= X_{2018} - (X_{2015}+dX)$
,~
$\mu_{y}= Y_{2018} - (Y_{2015}+dY)$.

\section{Selected sources}\label{sec:select}
Of the 494 sources referred to in Sect. \ref{sec:photometry}, 425 sources have been detected in the NIR (J and H) using SPHERE/IRDIS and in the optical using HST/WFPC2. 
To maximise robustness we excluded the sources  detected only in K and also the  76 sources that have fewer than three neighbours (within $R_{nbr}=50 pix$). 
This leaves 339 reliable sources that have been detected in the optical and in the NIR and that have at least three neighbours,  the minimum number of neighbours required to superimpose catalogues astrometrically.  

\begin{figure}
    \centering
    \includegraphics[trim=10 10 0 0,width=0.9\linewidth]{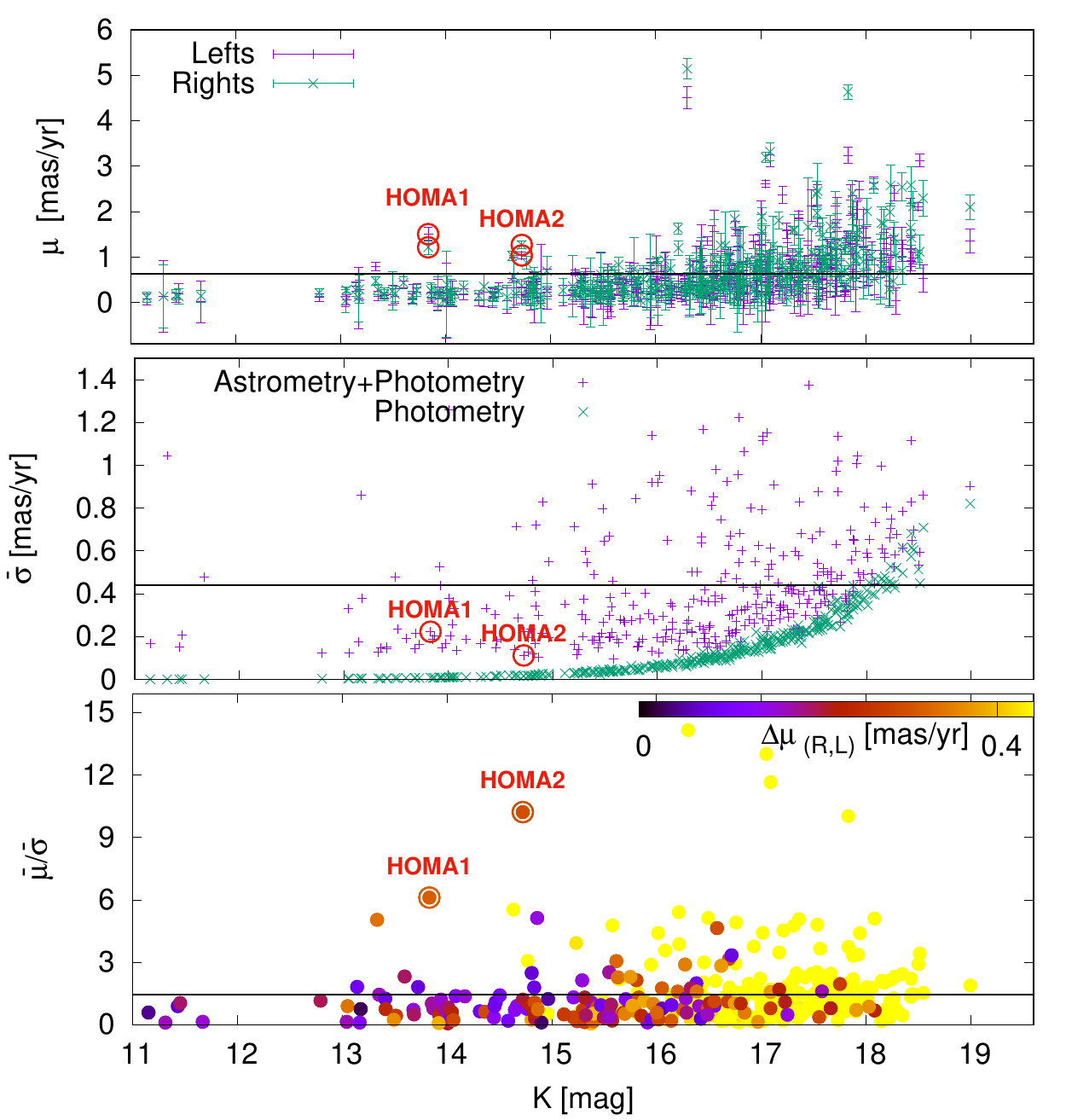}
    \caption{Astrometric analysis of 339 reliable sources, common between 2015 and 2018, left and right data.
    Top: Shift of stars, measured in the right (green crosses) and left (purple pluses) data vs their K magnitudes. 
    Middle: Total error (purple pluses) and photometric error (green crosses) on the positions of the detected sources, averaged between the left and right data vs their K magnitudes.
    Bottom: Shift of a star over its total error ($\frac{\mu}{\sigma}$) vs its K magnitudes, colour-coded by  $\Delta \mu _{(R,L)}$.
    The horizontal black lines show the average value for all 339 sources.}
    \label{fig:astrometry}
\end{figure}

The top plot of Fig. \ref{fig:astrometry} shows the shift ($\mu$) of the 339 reliable sources measured in the left and right data versus their K magnitude. 
The middle plot shows the error on the position of these sources versus their K magnitude. Errors on star positions come from the astrometric and photometric analysis (e.g. results from PSF fitting) on each data set. This plot shows that the error from astrometry is significantly larger than the photometry for the bright sources. For the faint stars the error on the photometry increases and becomes comparable with the astrometry error since the signal-to-noise ratio (S/N) decreases.
The bottom plot shows how large  the shift is compared to its error ($\frac{\mu}{\sigma}$). 
The colour bar indicates the difference between the shift of a star in the left and right data ($\Delta \mu _{(R,L)}$, Eq.~\ref{eq:shift}). 
All sources with  shifts inconsistent between left and right are yellow. For these sources, the measured shift between the left and right data is larger than the mean error of the reliable sources: 
\begin{equation}\label{eq:shift}
    \Delta \mu_{(R,L)}=\sqrt{(\mu _{x,L} - \mu_{x,R})^2+(\mu_{y,L} - \mu_{y,R})^2} 
.\end{equation}
When the detected sources become fainter, the astrometry measurements are less reliable due to the smaller S/N (the error on the star centres increase by inverse S/N), so that $\mu$ and $\Delta \mu _{(R,L)}$ increase with increasing stellar magnitudes.
For fainter sources this leads to larger errors in the positions so the measured shift will not be consistent between the left and right data. 
The mean shift of 339 reliable sources is 0.16 pix (0.64 mas/yr) and their mean total errors ($\bar{\sigma}$) is 0.11 pix (0.44 mas/yr). These values are shown in Fig. \ref{fig:astrometry} as a solid black horizontal line.

\begin{figure}
    \centering
    \includegraphics[trim=10 10 10 0,width=\linewidth]{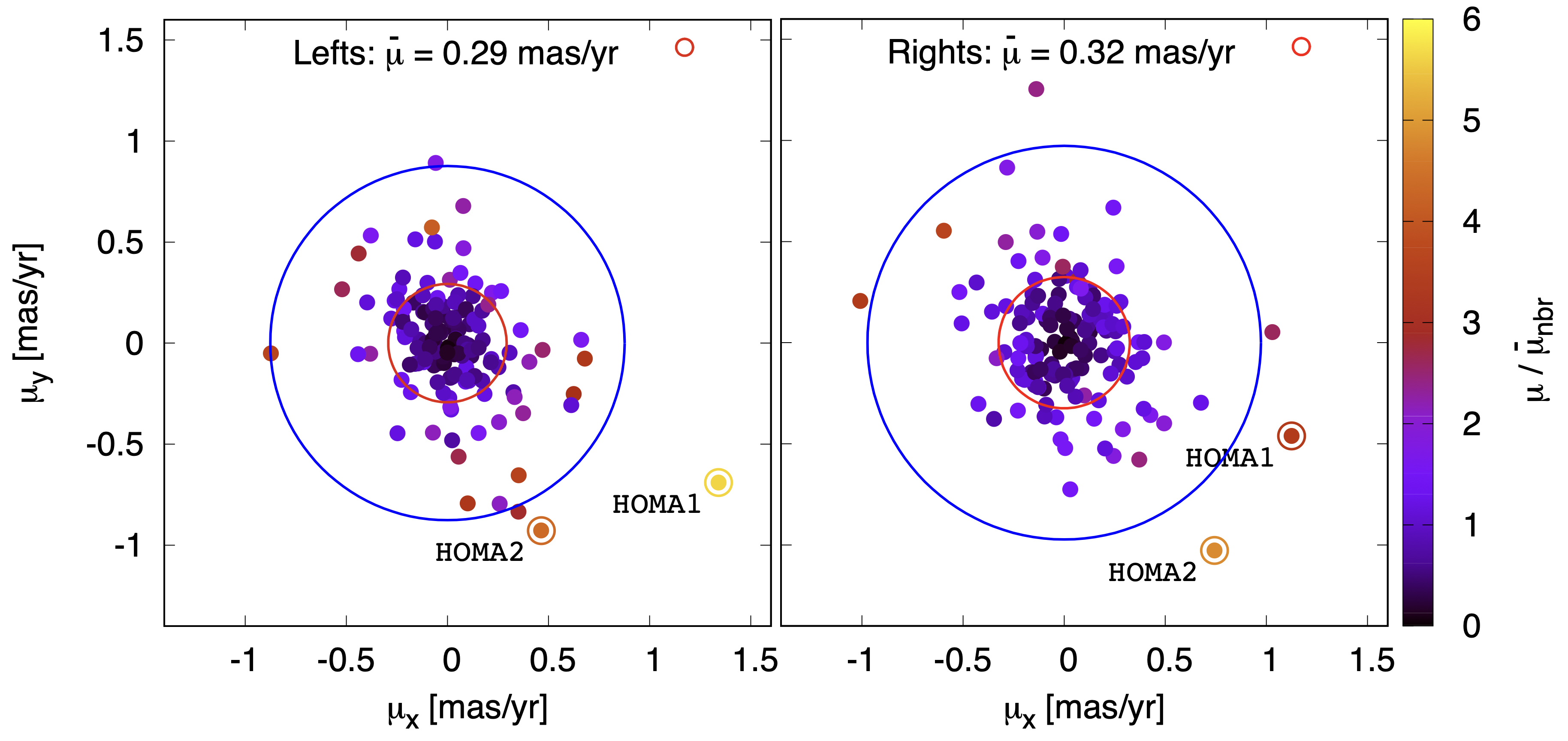}
    \caption{Estimated PM of RC sources from analysing the left (left) and right (right) data. 
    The radius of the large red and blue circle denotes  the mean shift of the 
RC sources  and  three times this value, respectively. 
    The colour-coding indicates the shift of each source compared to the mean shift of its neighbours ($\frac{\mu}{\bar{\mu}_{nbr}}$).
    }
    \label{fig:shiftXY}
\end{figure}

Of these 339 reliable sources, 141 have consistent velocities ($\Delta \mu_{(R,L)} \leq \bar{\sigma}$) between the  left and right data; they are known as reliable-consistent  (RC) sources.
Figure \ref{fig:shiftXY} shows the PM of RC sources measured from the analysis of the left and right data sets. The colour indicates the shift of each source compared to the mean shift of its neighbours ($\frac{\mu}{\bar{\mu}_{nbr}}$). 
The mean shift of RC sources in the left and right data are $0.07$\,pix ($0.29$\,mas/yr) and $0.08$\,pix ($0.32$\,mas/yr), respectively. 
The radius of the large red and blue circles is the mean shift of all RC sources and  three times this value, respectively. 
There are two sources with PMs that are three times larger than the mean shift of RC sources, and three times larger than the mean shift of their neighbours, in the left and the right data sets.
These two sources are identified as HOMAs (Fig. \ref{fig:shiftXY}); they are  both   brighter than 15~mag in K (16 in V) and have been detected with a high S/N in NIR filters and  in the visible. 

\begin{figure}
    \centering
    \includegraphics[trim=45 0 45 5,clip,width=0.9\linewidth]{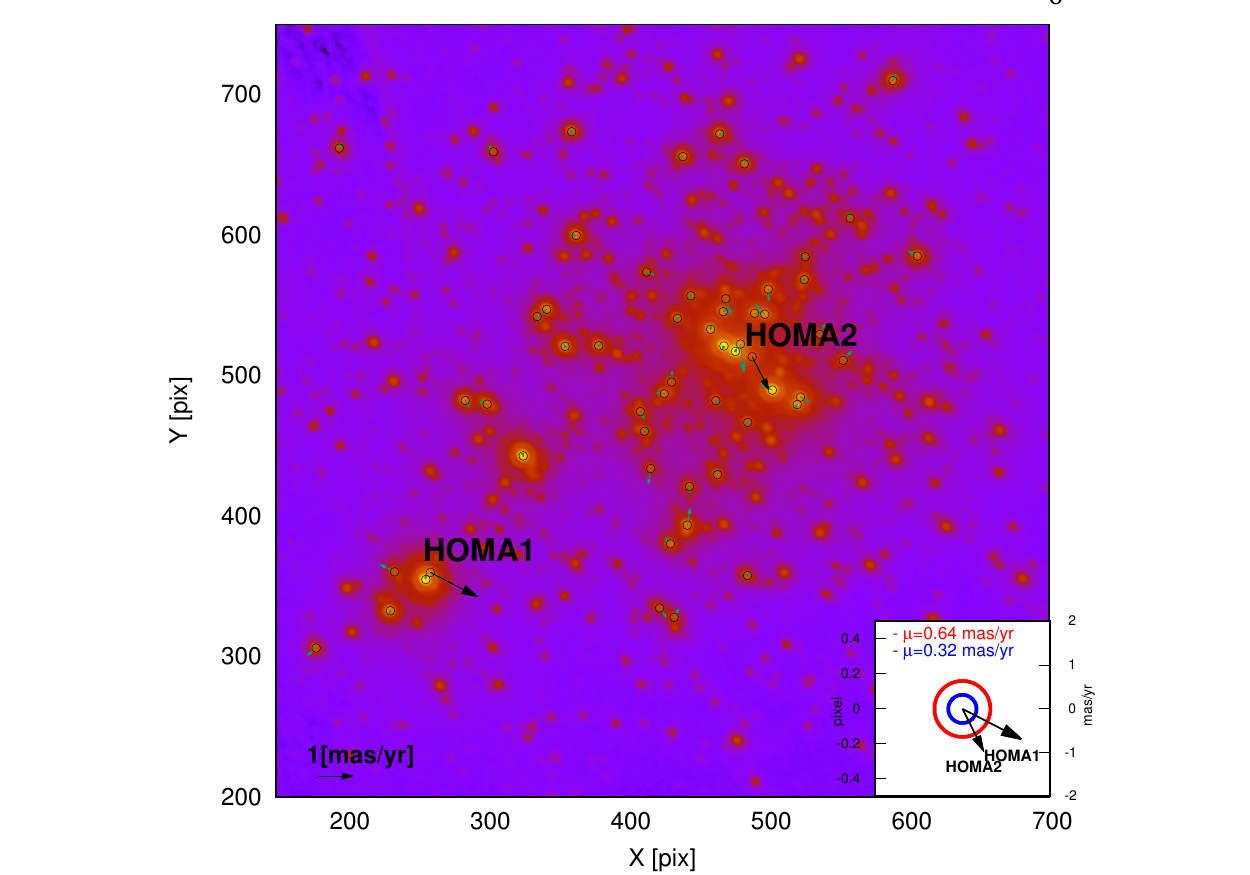}
    \caption{Image of R136 by IRDIS/K. Small black circles are bright reliable sources (K $<$ 15 mag). The green and black arrows indicate the PM of these sources, magnified 100 times.
    The inset in the bottom right corner shows 1 pixel size (1pix=4mas/yr) and the amount of the HOMAs shifts compared to the mean shift of 339 reliable sources ($\bar{\mu}=0.64$\,mas/yr) and 141 RC sources ($\bar{\mu}=0.32$\,mas/yr) respectively as a large red and blue circle.}
    \label{fig:homas}
\end{figure}

These stars are shown in Fig. \ref{fig:homas}, together with  62 bright sources ($K\leq15$) for comparison. 
Notably, both of the  HOMAs are located very close to R136a1 (separation of 155~mas) and R136c (separation of 72~mas), which are bright K-band sources \footnote{See Fig.~\ref{fig:homaszoom} to compare these separations (72 and 155\,mas) with the radius of the FWHM (31.81\,mas), 
and the first dark (37.90\,mas) and first bright (50.9\,mas6) rings of the Airy Disk.}.
The analysis of the X-ray spectra of R136c shows the high X-ray hardness and luminosity in addition to the higher plasma temperature than typical for single early-type stars, suggesting that  it is most likely to be a colliding-wind binary system \citep{zwart2002,townsley2006,guerrero2008}.

Table \ref{table:infoRun} gives the IVUJHK magnitudes and PMs of the two HOMAs and the five brightest sources in the FOV. 
HOMA1 and HOMA2 have PMs of $1.36\pm0.22$~mas/yr and $1.15\pm0.11$~mas/yr, respectively. For the distance modulus of 18.49 \citep{pietrzyski2013,gibson2000} 1~mas/yr is 236~km/s, these HOMAs have a tangential velocity of $321\pm52$~km/s and $273\pm26$~km/s.

\begin{table*}
\small
\caption{Magnitudes and PMs of the HOMAs and the five brightest sources in the core of R136.
H95 and K21 are the identification of these sources from  Hunter et al. (1995) and Khorrami et al (2021), respectively.
K$_{2018}$, K$_{2015}$, H, and J are the magnitudes in the SPHERE/IRDIS filters \citep{khorrami2017,khorrami2021}.
I, V, and U are the magnitudes in the HST F814W, F555W, and F336W filters \citep{hunter95}.
$\mu_{x}$ and $\mu_{y}$ are the 2D PMs estimated from their position shift within 3.06 years in the left and right data.
}
\centering
\begin{tabular}{|l |c c c |c c c |cc|cc|}  
\hline
Name&K$_{2018}$, K$_{2015}$& H& J& I & V & U &$\mu_x$[mas/yr] &$\mu_y$ [mas/yr]&$\mu_x$ [mas/yr]&$\mu_y$[mas/yr] \\
(H95/K21)&\multicolumn{3}{|c|}{(SPHERE/IRDIS)}&\multicolumn{3}{|c|}{(HST/WFPC2)}& \multicolumn{2}{c}{(left)}&\multicolumn{2}{|c|}{(right)}\\
\hline
HOMA1 (124/22)&13.83, 13.84&14.41&14.33&15.46&15.97&-&$1.34\pm0.12$& $-0.69\pm0.09$&$1.12\pm0.13$& $-0.46\pm0.10$\\
HOMA2 (107/51)&14.72, 14.49&14.68&14.77&15.34&15.74&14.58&$0.46\pm0.05$&$-0.93\pm0.05$&$0.74\pm0.04$&$-1.03\pm0.08$\\
\hline

R136a1 (3/1)&11.15, 11.07&11.29&11.33&12.18&12.84&11.56&$0.01\pm0.08$&$0.07\pm0.09$&$0.05\pm0.08$&$0.11\pm0.08$\\
R136a2 (5/3)&11.43, 11.32&11.70&11.55&12.84&12.96&11.94&$-0.01\pm0.09$&$-0.10\pm0.05$&$0.04\pm0.07$&$-0.16\pm 0.08$\\
R136a3 (6/4)&11.45, 11.44& 11.73& 11.77 &12.46& 13.01& 11.86&$-0.07\pm0.05$&$0.12\pm0.10$&$0.09\pm0.10$&$0.11\pm0.14$\\
R136c (10/2)&11.31, 11.48& 11.98& 11.78& 12.71& 13.47& 12.52&$-0.00\pm0.70$&$0.14\pm0.35$&$0.13\pm0.63$&$0.05\pm  0.27$\\
R136b (9/5)&11.67, 11.66& 11.89& 11.67& 12.76& 13.32& 12.29& $0.00\pm0.37$& $0.02\pm0.28$&$0.12\pm0.10$&$0.08\pm  0.08$\\
\hline
\end{tabular} 
\label{table:infoRun}  
\end{table*}

\section{Conclusion}\label{sec:conclusion}
We obtained VLT/SPHERE/IRDIS K-band images of the core of R136 at two epochs, separated by 3.06 years.
A careful scrutiny of two-epoch astrometry enabled us to discard instrumentally apparent moving sources.
We found 494 suitable sources in common between these four images in the K band, where 339 sources have also been detected in IRDIS-J and -H, and in the optical by HST/WFPC2.
Comparing the 2D shift of stars in left and right data sets, only 141 sources have consistent PMs between these data sets.
Among these RC sources, two stars have a PM that is three times larger than the average value ($0.29$\,mas/yr, left, and $0.32$\,mas/yr, right) and their neighbours. 
According to our analysis, these two bright sources (see Table \ref{table:infoRun}) are potential HOMAs. 
These stars have shifted more than the mean shift of all reliable sources (Fig. \ref{fig:astrometry}-top), three times that of the average shift of all RC sources and their neighbours (Fig. \ref{fig:shiftXY}), and six times their astrometric errors (Fig. \ref{fig:astrometry}-bottom), with a tangential velocity of $321\pm52$ km/s and $273\pm26$ km/s. 
The stars could have acquired such high velocities through a number of mechanisms, none of which are considered probable: (i) during the assembly of the cluster (e.g. VFTS16 currently very distant from R136, \citealt{lennon2018}), (ii) following the supernova explosion of a binary companion (too early for R136 given its youth, \citealt{renzo2019}); (iii) via three-body interactions with a massive binary (possible, though incredibly rare); (iv) an encounter with a compact massive body (although there is no evidence for an intermediate mass black hole in R136). Alternatively, they could be slower moving foreground sources (though the presence of two sources within the small IRDIS FOV is very unlikely).
Comparing the location of HOMAs in the CMDs (Fig.~\ref{fig:cmds}), HOMA2 is closer to the main sequence than HOMA1. 
HOMA1(HOMA2) has V-I of 0.51(0.40)mag and J-K of 0.49(0.28)mag, making them unlikely to be foreground Galactic stars.
If HOMA1 is a foreground galactic star, then its tangential velocity is $\sim108$~km/s at a distance of 20 Kpc (edge of the Milky Way toward the LMC).
Since the localisation of these sources is challenging given their proximity to R136a1 and R136c, future observations are necessary to confirm their high PMs and status as HOMAs. These sources are   expected to have high radial velocities, so they are the desired targets for the spectroscopic observations of the core of R136 (e.g. MUSE/NFM \citealt{Castro21}). 
Pending future gain in spatial resolution by extremely large telescopes, the present work is also the first step towards constraints on the kinematics of other galaxies than our Milky Way.

\begin{acknowledgements}
We thank our reviewer, C. Evans, for his constructive comments which improved this letter. The SFM project has received funding from the European Union's Horizon 2020 research and innovation program under grant agreement No 687528. ZK acknowledges the support of a STFC Consolidated Grant (ST/K00926/1).
This work has made use of the SPHERE Data Centre, jointly operated by OSUG/IPAG (Grenoble), PYTHEAS/LAM/CeSAM (Marseille), OCA/Lagrange (Nice), Observatoire de Paris/LESIA (Paris), and Observatoire de Lyon (OSUL/CRAL).
This work was supported by the "Programme National de Physique Stellaire" (PNPS) of CNRS/INSU co-funded by CEA and CNES.
ASMB is funded by the European Research Council H2020-EU.1.1 ICYBOB project (Grant No. 818940). JP and RW acknowledge support by the Czech Science Foundation project no. 19-15008S and by the institutional project RVO:67985815.

\end{acknowledgements}

%

\begin{thebibliography}{}
\bibitem[\protect\citeauthoryear{Anderson \& van der Marel}{2010}]{lct2010} Anderson J., van der Marel R.~P., 2010, ApJ, 710, 1032. 
\bibitem[\protect\citeauthoryear{Andersen et al.} {2009}]{andersen2009} Andersen, M., Zinnecker, H., Moneti, A., et al.\ 2009, ApJ, 707, 1347 
\bibitem[\protect\citeauthoryear{Bestenlehner et al.}{2020}]{Bestenlehner2020} Bestenlehner J.~M., Crowther P.~A., Caballero-Nieves S.~M., et al., 2020, MNRAS.tmp, doi:10.1093/mnras/staa2801
\bibitem[\protect\citeauthoryear{Beuzit et al.}{2019}]{sphere} Beuzit, J. L., Vigan, A., Mouillet, D., et al., 2019, A\&A vol. 631, id.A155, 36
\bibitem[\protect\citeauthoryear{Campbell et al.}{2010}]{campbell2010} Campbell, M.A., Evans, C.J., Mackey, A.D., Gieles, M., Alves, J., Ascenso, J., Bastian, N., Longmore, A.J. 2010, MNRAS 405, 421
\bibitem[\protect\citeauthoryear{Castro et al.}{2021}]{Castro21} Castro, N., Roth, M. M., Weilbacher, P. M., et al. 2021, arXiv:2102.01113
\bibitem[\protect\citeauthoryear{Cignoni et al.}{2015}]{Cignoni2015} Cignoni M., Sabbi E., van der Marel R.~P., et al., 2015, ApJ, 811, 76.
\bibitem[\protect\citeauthoryear{Crowther et~al.}{2010}]{crowther2010} Crowther, P. A., Schnurr, O., Hirschi, R., et al., 2010, MNRAS, 408, 731
\bibitem[\protect\citeauthoryear{Crowther et al.}{2016}]{crowther2016} Crowther, P.~A., Caballero-Nieves, S.~M., Bostroem, K.~A., et al.,2016, MNRAS, 458, 624
\bibitem[\protect\citeauthoryear{Crowther}{2019}]{crowther2019} Crowther, P.~A., 2019, Galaxies, 7, 88, doi:10.3390/galaxies7040088
\bibitem[\protect\citeauthoryear{de Koter, Heap, \& Hubeny}{1998}]{dekoter1998} de Koter A., Heap S.~R., Hubeny I., 1998, ApJ, 509, 879. doi:10.1086/306503
\bibitem[\protect\citeauthoryear{De Marchi et al.}{2011}]{demarchi2011} De Marchi G., Paresce F., Panagia N., et al., 2011, ApJ, 739, 27.
\bibitem[\protect\citeauthoryear{Evans et al.}{2011}]{evans2011} Evans C.~J., et al., 2011, A\&A, 530, A108
\bibitem[\protect\citeauthoryear{Gibson}{2000}]{gibson2000} Gibson, B.~K., 2000, MemSAI, 71, 693 
\bibitem[\protect\citeauthoryear{Guerrero \& Chu}{2008}]{guerrero2008} Guerrero M.~A., Chu Y.-H., 2008, ApJS, 177, 216
\bibitem[\protect\citeauthoryear{Hunter et al.}{1995}]{hunter95} Hunter, D. A., Shaya, E. J., Holtzman, J. A. 1995,ApJ 448, 179
\bibitem[\protect\citeauthoryear{Khorrami et al.}{2017}]{khorrami2017} Khorrami, Z., Vakili, F., Lanz, T., et al., 2017, A\&A, 602, A56
\bibitem[\protect\citeauthoryear{Khorrami et al.}{2021}]{khorrami2021} Khorrami Z., Langlois M., Clark P.~C., Vakili F., Buckner A.~S.~M., Gonzalez M., Crowther P., et al., 2021, MNRAS, 503, 292. doi:10.1093/mnras/stab388
\bibitem[\protect\citeauthoryear{Langlois et al.}{2014}]{maud14} Langlois, M., Vigan, A., Dohlen, K., et al.\ 2014, SPIE, 9147, 91479P
\bibitem[\protect\citeauthoryear{Lennon et al.}{2018}]{lennon2018} Lennon, D.~J., Evans, C. J., van der Marel, R. P. et al. 2018, A\&A, 619, A78
\bibitem[\protect\citeauthoryear{Maire et al.}{2016}]{maire} Maire A.-L., Langlois M., Dohlen K., Lagrange A.-M., Gratton R., Chauvin G., Desidera S., et al., 2016, SPIE, 9908, 990834. doi:10.1117/12.2233013
\bibitem[\protect\citeauthoryear{Platais et al.}{2015}]{platais2015} Platais, I., van der Marel, R. P., Lennon, D. J., et al. 2015, AJ, 150:89
\bibitem[\protect\citeauthoryear{Platais et al.}{2018}]{platais2018} Platais, I., Lennon, D. J., van der Marel, R. P., et al. 2018, AJ, 156:98
\bibitem[\protect\citeauthoryear{Pietrzy{\'n}ski et al.}{2013}]{pietrzyski2013} Pietrzy{\'n}ski, G., Graczyk, D., Gieren, W., et al.\ 2013, NAT, 495, 76 
\bibitem[\protect\citeauthoryear{Renzo et al.}{2019}]{renzo2019} Renzo M., et al., 2019, A\&A, 624, A66
\bibitem[\protect\citeauthoryear{Portegies Zwart, Pooley, \& Lewin}{2002}]{zwart2002} Portegies Zwart S.~F., Pooley D., Lewin W.~H.~G., 2002, ApJ, 574, 762. 
\bibitem[\protect\citeauthoryear{Townsley et al.}{2006}]{townsley2006} Townsley L.~K., Broos P.~S., Feigelson E.~D., et al., 2006, AJ, 131, 2164
\end{thebibliography}
%

\appendix
\section{Extra figures}
\begin{figure}
    \centering
    \includegraphics[trim=120 0 120 0,clip,width=0.3\linewidth]{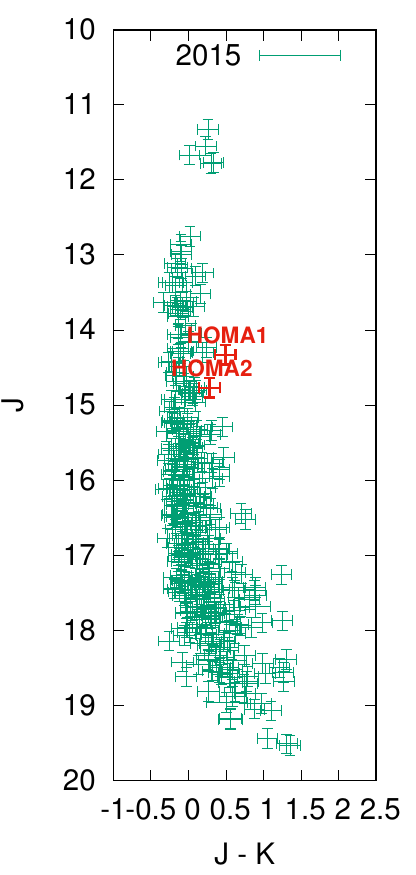}
     \includegraphics[trim=120 0 120 0,clip,width=0.3\linewidth]{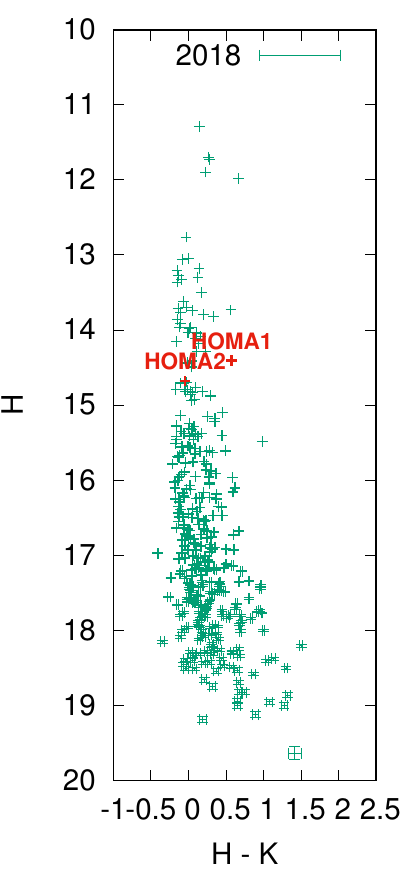}
    \includegraphics[trim=120 0 120 0,clip,width=0.3\linewidth]{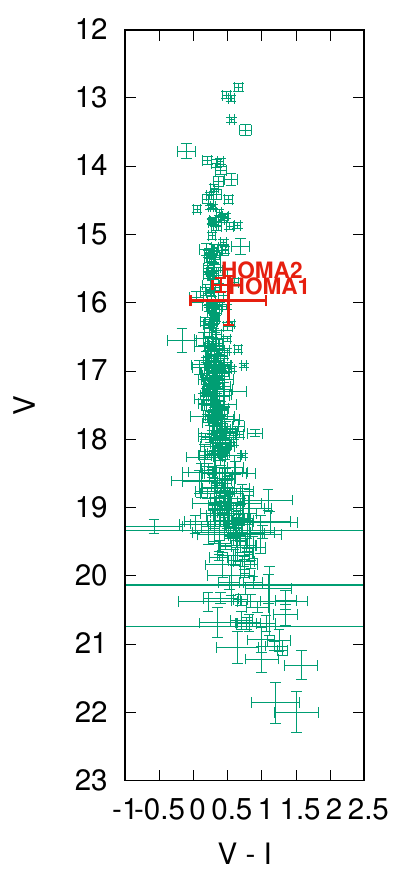}
       \caption{Colour--magnitude diagram 
       of 339 reliable sources. 
       The data are   from the SPHERE/IRDIS catalogue in the NIR (J, H, K) and HST/WFPC2 data in the visible (V and I)
       CMD in J-K (left), H-K (middle), and   V-I (right).
       The red symbols shows the HOMAs.}
    \label{fig:cmds}
\end{figure}

\begin{figure}
    \centering
    \includegraphics[trim=60 0 45 10,width=\linewidth]{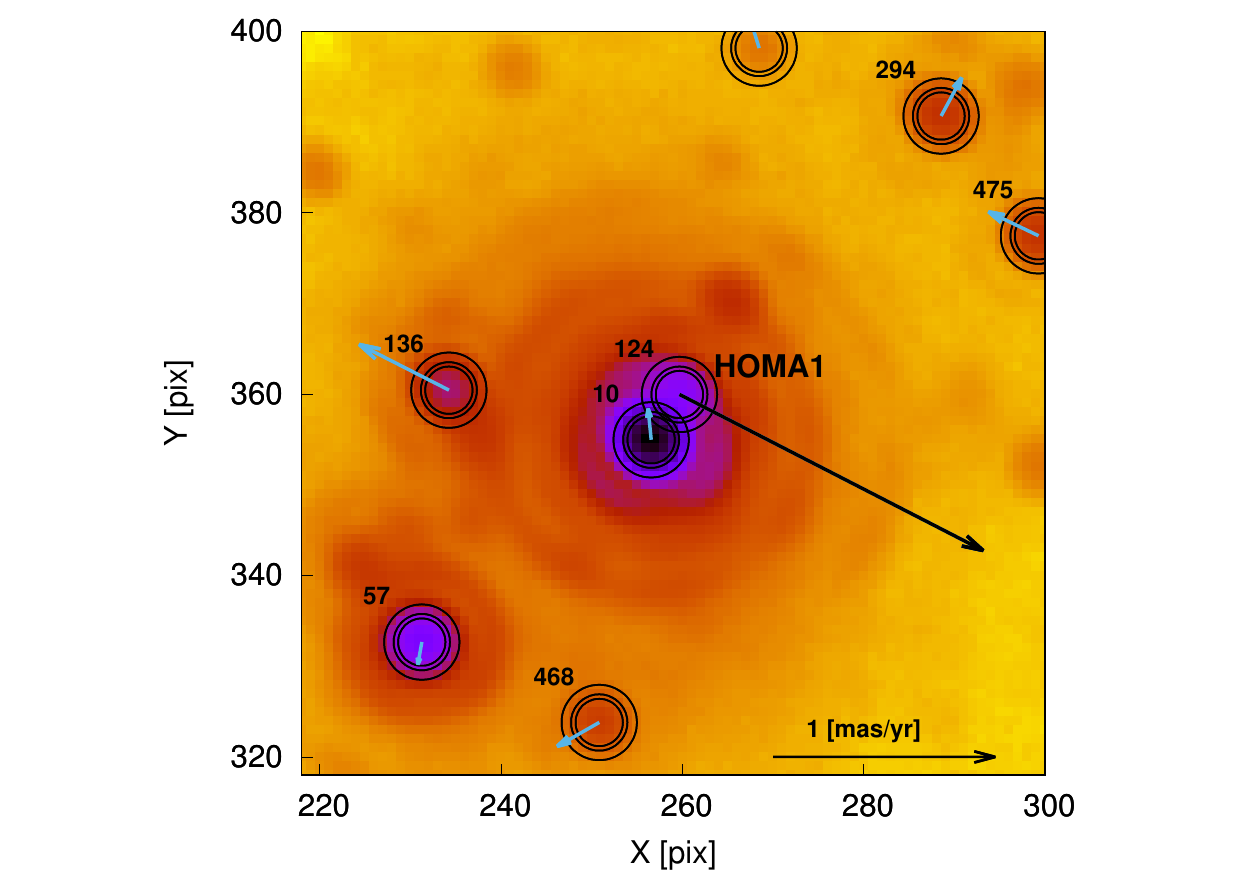}\\
    \includegraphics[trim=60 0 45 10,width=\linewidth]{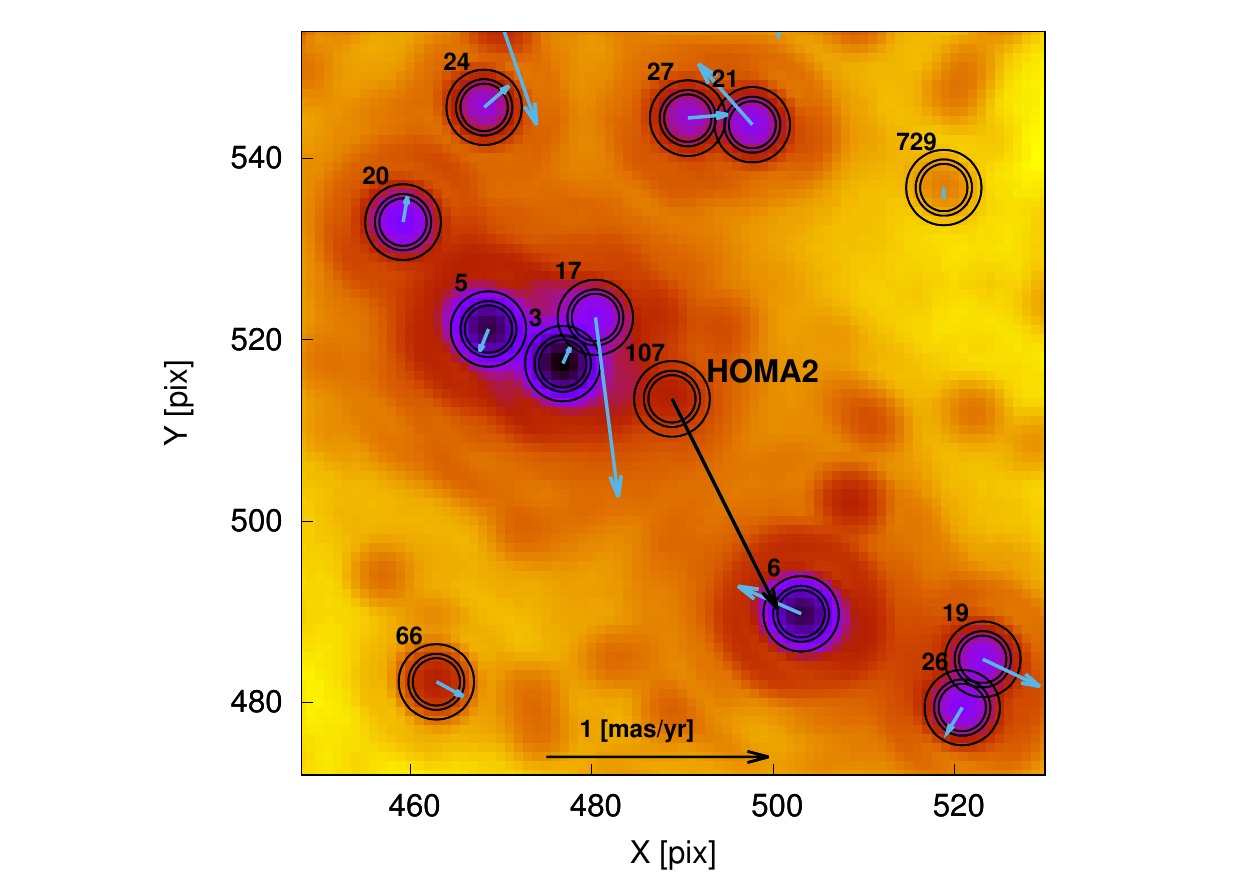}
    \caption{Same as Fig.~\ref{fig:homas}, but zoomed in for a FOV of $1"\times1"$ , centred on HOMA1 (top) and HOMA2 (bottom). 
    The diameter of three black circles around each bright reliable sources (K $<$ 15 mag) are the FWHM, 
    and the first dark and first bright rings of the Airy Disk.
    The ID number  of each source (H95, from Hunter et al. 1995) is shown (above and to the left).
    The green and black arrows indicate the PM of these sources, magnified 100 times.
    }
    \label{fig:homaszoom}
\end{figure}

\end{document}